\documentclass[english,nofootinbib, aps, floatfix, superscriptaddress, prd, twocolumn,longbibliography]{revtex4-1}
\usepackage[T1]{fontenc}
\usepackage{graphicx}
\usepackage[T1]{fontenc}
\usepackage[latin9]{inputenc}
\usepackage{color}
\usepackage{babel}
\usepackage{float}
\usepackage{amsmath}
\usepackage{amssymb}
\usepackage{physics}
\usepackage{slashed}
\usepackage[unicode=true, bookmarks=false, breaklinks=false,pdfborder={0 0 1},colorlinks=true]{hyperref}
\usepackage{hyperref,overpic}
\hypersetup{
    colorlinks=true,
    linkcolor=blue,
    filecolor=magenta,      
    urlcolor=blue,
    citecolor=magenta,
}

\makeatletter

\providecommand{\tabularnewline}{\\}
\usepackage{tikz}
\usetikzlibrary{calc}

\bibpunct{[}{]}{;}{n}{}{}
\usepackage{amsfonts}

\DeclareMathOperator{\sgn}{sgn}

\makeatother

\begin{document}
\include{macros}

\title{Conformal field theories in a magnetic field}

\author{Rufus Boyack}
\affiliation{Department of Physics and Astronomy, Dartmouth College, Hanover, NH 03755, USA}
\affiliation{D\'epartement de physique, Universit\'e de Montr\'eal, Montr\'eal (Qu\'ebec), H3C 3J7, Canada}

\author{Luca Delacr\'etaz}
\affiliation{Kadanoff Center for Theoretical Physics, University of Chicago, Chicago, IL 60637, USA}
\affiliation{James Franck Institute, University of Chicago, Chicago, IL 60637, USA}

\author{\'Eric Dupuis}
\affiliation{D\'epartement de physique, Universit\'e de Montr\'eal, Montr\'eal (Qu\'ebec), H3C 3J7, Canada}

\author{William Witczak-Krempa}
\affiliation{D\'epartement de physique, Universit\'e de Montr\'eal, Montr\'eal (Qu\'ebec), H3C 3J7, Canada}
\affiliation{Institut Courtois, Universit\'e de Montr\'eal, Montr\'eal (Qu\'ebec), H2V 0B3, Canada}
\affiliation{Centre de Recherches Math\'ematiques, Universit\'e de Montr\'eal; P.O. Box 6128, Centre-ville Station; Montr\'eal (Qu\'ebec), H3C 3J7, Canada}

\begin{abstract}
We study the properties of 2+1d conformal field theories (CFTs) in a background magnetic field. Using generalized particle-vortex duality, we argue that in many cases of interest the theory becomes gapped, which allows us to make a number of predictions for the magnetic response, background monopole operators, and more. Explicit calculations at large $N$ for Wilson-Fisher and Gross-Neveu CFTs support our claim, and yield the spectrum of background (defect) monopole operators. Finally, we point out that other possibilities exist: certain CFTs can become metallic in a magnetic field.
Such a scenario occurs for a Dirac fermion coupled to a Chern-Simons gauge field, where a non-Fermi liquid is argued to emerge.

\end{abstract}

\maketitle
\tableofcontents{}

\section{Introduction}

\begin{figure}
\centerline{
\begin{overpic}[width=0.7\linewidth,tics=10]{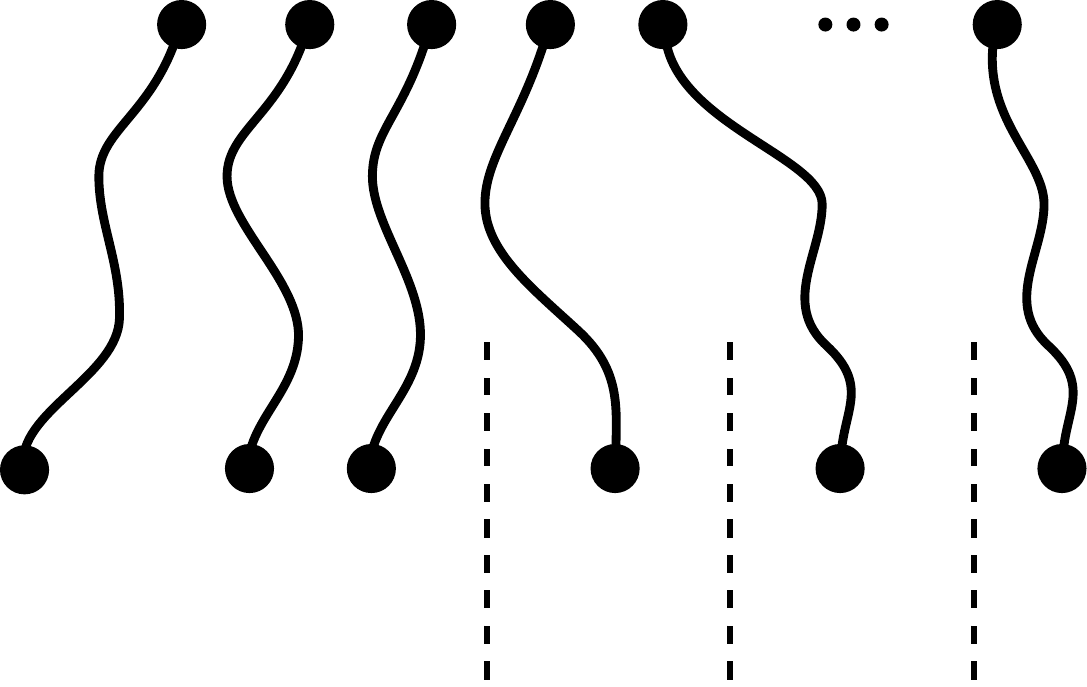}
	 \put (15,65) {\rotatebox{45}{CFT${}_1$}} 
	 \put (27,65) {\rotatebox{45}{CFT${}_2$}} 
	 \put (39,65) {\rotatebox{45}{CFT${}_3$}} 
	 \put (55,68) {\large$\cdots$} 
	 \put (-10,45) {\large $+B$} 
	 \put (10,9) { gapped} 
	 \put (48,9) { metal} 
	 \put (71,9) { NFL} 
	 \put (95,9) { ?} 
\end{overpic} 
}
\caption{\label{fig_phases}Landscape of phases that emerge from deforming a 2+1d CFT with a magnetic field $B$. CFTs leading to gapped phases can be distinguished by their Wilsonian coefficients in the EFT \eqref{eq:logZ}.}
\end{figure}
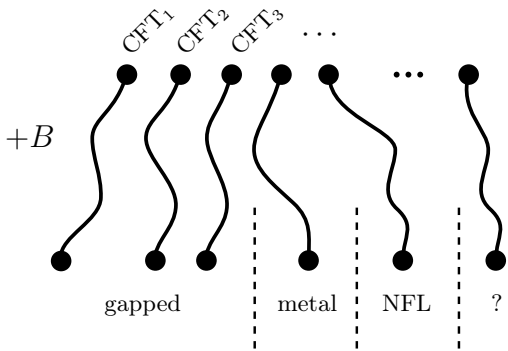

The application of an external magnetic field can give rise to entirely distinct states of quantum matter. For example, an electron gas in two dimensions (2DEG) is a conventional metal, while a strong magnetic field can transform it into a gapped topological state with emergent anyons and robust edge states, or a Wigner crystal. 
But what if the initial state is not as simple as a regular metal? Here, we study this question in the context of an important family of gapless quantum states: conformal field theories. These quantum critical states can describe a quantum phase transition such as the insulator-to-superfluid transition in the celebrated XY model, or a stable phase of matter such as a Dirac semimetal, or more interestingly quantum electrodynamics (QED) with massless Dirac fermions, as could arise in frustrated quantum magnets. In this context, the general question we address can be formulated as follows: what is the landscape of phases that can be obtained by deforming 2+1d CFTs with a magnetic field $B$? 
One may expect that a CFT placed in a magnetic field always develops a gap. This is indeed what happens in free theories, where the magnetic field produces Landau levels. We will argue below, using S-duality, that this is a generic possibility for interacting CFTs as well. However, we also find examples of CFTs where a magnetic field produces dramatically different phases, including metals and non-Fermi liquids. Figure~\ref{fig_phases} illustrates part of this landscape.

Mapping out the phases accessible from a CFT by turning on a magnetic field has interesting parallels with those obtained instead by turning on a chemical potential $\mu$. In this case, the generic expectation is that the CFT enters a superfluid phase \cite{Feynman1953,leggett1973topics,sachdev_compressible_2012,Joyce2022}. This issue was recently revisited in the study of local operators of large charge $Q$ \cite{hellerman_on_2015,monin_semiclassics_2017,Jafferis2018}. Indeed, the state-operator correspondence of CFTs provides a map between the spectrum of states on the sphere and local operators of the CFT -- large charge operators therefore map to finite-density phases on the sphere in the thermodynamic limit. Effective field theories (EFTs) for these phases then allow for controlled descriptions of sectors of otherwise strongly coupled CFTs. 
More generally, certain aspects of the spectra of CFTs become tractable at large quantum numbers, including large charge $Q$ \cite{hellerman_on_2015,monin_semiclassics_2017,Jafferis2018,Dondi2023}, large spin $J$ \cite{Fitzpatrick2013,Komargodski2012,CaronHuot2017}, and large scaling dimension $\Delta$ \cite{Lashkari2018,Delacretaz:2020nit,Benjamin:2023qsc}, even in the absence of an underlying control parameter in the CFT such as large $N$ or weak coupling. Our analysis folds into this line of research, but with an unusual twist: 
the large number here is the background magnetic flux $Q_B = \int_{S^2}\frac{B}{4\pi}\in \frac12 \mathbb Z$  piercing the sphere, so that these states do not map to local operators of the CFT but rather a class of defect operators called {\em background  monopoles}~\cite{kapustin_generalized_2011,sachdev_compressible_2012}. While somewhat less familiar, these are also part of the universal data characterizing 2+1d CFTs with a U(1) symmetry. Their two-point functions have power-law decay, whose exponent $\Delta$ labels their scaling dimension. At large $Q_B$, we will see that dimensional analysis (and the assumption that the CFT has a finite magnetic susceptibility) implies that the dimension of the lightest background monopole scales as $\Delta\sim Q_B^{3/2}$. Furthermore, the assumption that the phase is gapped leads to an effective action that predicts a series of corrections in integer powers of $1/Q_B$: $\Delta\sim Q_B^{3/2} + Q_B^{1/2} + O(Q_B^{-1/2})$. This is in contrast to the regular large-charge expansion~\citep{hellerman_on_2015,monin_semiclassics_2017,cuomo2020large}, where gapless IR fluctuations typically lead to an $O(Q^0)$ piece.

The rest of this paper can be summarized as follows: in Sec.~\ref{sec_EFT}, we consider CFTs that become gapped when placed in a magnetic field. The gap allows one to integrate out matter to obtain a local, gauge-invariant effective action for the background field $A_\mu$, whose Wilsonian coefficients characterize the CFT in this phase. For example, these coefficients include the uniform magnetic susceptibility of the CFT, as well as its finite-wavevector corrections. In Sec.~\ref{sec_Higgs}, we give an argument for the genericity of the gapped phase by passing to the vortex dual (or S-dual) description: if the S-dual CFT becomes a superfluid when placed in a chemical potential, then the original CFT will be in a Higgs phase when placed in a background magnetic field. This argument assumes certain gauge fluctuations are suppressed, and therefore it applies only to weakly coupled CFTs, but we expect it to be indicative of fairly generic behavior. Section~\ref{sec:FreeEnergyCalcs} is devoted to a detailed analysis of background monopole scaling dimensions in large-$N$ CFTs. We confirm the absence of an $O(Q_B^0)$ term in all the models studied. 
Finally, in Sec.~\ref{sec:nfl} we identify scenarios where a magnetic field converts a CFT into a metal or even a non-Fermi-liquid. This strongly correlated behavior at low energies arises even though the original CFT is weakly coupled. Another alternative to gapped phases is provided by holographic CFTs, which are expected to become extremal black holes when placed in a magnetic field.

CFTs placed in a magnetic field have of course been studied in the past, in particular in the context of weakly coupled 
\cite{klimenko_3dGN_1991,Klimenko1992,krive_dynamical_1992,gusynin_dynamical_1995,gusynin_largeNQED_2003,scherer_renormalization_2012,lenz_magnetic_2023}
or holographic \cite{hartnoll_hall_2007,Hartnoll2007} theories. However, to our knowledge, they have not been approached from the perspective of the effective description that emerges in a magnetic field.

\section{CFT in an external magnetic field}\label{sec_EFT}

\subsection{CFT in flat spacetime}

Let us first begin by considering a CFT in flat space with a global U(1) symmetry, and corresponding conserved current density operator $J_\mu$. In this subsection, we shall further assume that the CFT preserves parity; the case of parity-violating CFTs will be discussed in Secs.~\ref{sec:functional} and \ref{sec:nfl}. We now add an external uniform magnetic field $B$, which in 2+1d is a real number that can take both signs. It is implemented by coupling the theory to a background gauge field $A_\mu$ and adding a relevant term to the action $\int_x A_\mu J^\mu$. Before discussing the possible fates of the system, let us examine some basic properties. The magnetic field will induce a non-zero energy density in the system:
\begin{equation} 
\label{eq:energy}
\varepsilon = \frac{4}{3}\chi_m |B|^{3/2}.
\end{equation}
In the ground state at zero field, the energy density vanishes by conformal invariance, and the power 3/2 follows since the magnetic field has units of $1/(\mbox{length})^2$. We note that a term that is odd in $B$, $|B|^{3/2}\sgn B$, cannot be included since it is odd under time-reversal (and parity) whereas the energy density of the CFT is even. The dimensionless coefficient $\chi_m$ is the \emph{magnetic susceptibility} of the CFT. It is a CFT-dependent coefficient that quantifies to what extent the system responds to an external magnetic field. It will grow with the number of charged degrees of freedom. For instance, for $N$ massless free Dirac fermions or complex bosons coupled symmetrically to a $B$ field, $\chi_m$ will be linear in $N$. From the above equation, we can obtain the magnetization by differentiating \eqref{eq:energy} with respect to $B$:
\begin{equation}
M=\frac{d\varepsilon}{dB} = 2\chi_m |B|^{1/2}\sgn B
\end{equation}
so that the magnetization will change sign as $B$ goes from positive to negative. The susceptibility is the derivative of $M$ with respect to $B$, up to an overall power of $B$ to make it a dimensionless property of the CFT:
\begin{equation}
\chi_m = |B|^{1/2}\frac{dM}{dB}= |B|^{1/2}\frac{d^2\varepsilon}{dB^2}. 
\end{equation}

Now, there are two basic possibilities for the CFT in a magnetic field: (1) the system becomes gapped or (2) it remains gapless. In the simplest cases of free CFTs, a gap will appear. Indeed, for free Dirac fermions, dispersionless relativistic Landau levels (LLs) are formed: $E_n= \sgn(n) \sqrt{2|nB|}$, $n=0,\pm 1, \pm 2, \dots$. One entirely fills the negative LLs, while the $n=0$ LL is half-filled. In order to excite the system, one has to pay an energy cost $|2B|^{1/2}$ to promote a fermion to the $n=1$ LL; we thus have a gapped theory. Nonetheless, the system possesses an infinite degeneracy due to the different ways of half-filling the $n=0$ LL. We note that an exact degeneracy remains even when working on a finite sphere pierced by a flux, since the Atiyah-Singer index theorem protects the zero modes. Similarly, a free complex boson CFT perturbed by a magnetic field will also possess a gap since the vacuum has no boson, but adding a single boson will cost a minimal energy $\propto |B|^{1/2}$. There is no ground-state degeneracy in the bosonic case.

Before we analyze the justification for the gapped scenario in more generality, let us now examine the implications of having a field-induced gap.

\subsection{Free-energy functional in gapped phases} \label{sec:functional}

All degrees of freedom can be integrated out to obtain a generating functional $\log Z[A_\mu,g_{\mu\nu}]$; assuming the phase is gapped, this functional is local, with higher derivative terms suppressed by the gap of the phase. Here, $A_\mu$ is a general spacetime-dependent vector field, thus more general than the gauge field that would give rise to a constant magnetic field.
See Ref.~\cite{Golkar2014} for such a construction in the context of relativistic QH systems. Requiring the generating functional to be gauge invariant and diffeomorphism invariant, as well as invariant under Weyl transformations $g_{\mu\nu}(x)\to \Omega(x)^2 g_{\mu\nu}(x)$, one obtains (see App.~\ref{app_gapped})
\begin{widetext}
\begin{equation}
\label{eq:logZ}
\begin{split}
\log Z[A,g]
	&= \frac{\nu}{4\pi} \int d^3 x \, \epsilon^{\mu\nu\lambda} A_\mu \partial_\nu A_\lambda \\
	&\quad+ \frac43 \chi_m \int d^3 x \sqrt{g} F^{3/2}\\
	&\quad+ \zeta \int d^3 x \, F \epsilon^{\mu\nu\lambda} u_\mu \partial_\nu u_\lambda
	+ \frac{\kappa}{8\pi} \int d^3 x \,\epsilon^{\mu\nu\lambda}\epsilon_{\alpha\beta\gamma} A_\mu u^\alpha \left(\nabla_\nu u^\beta \nabla_\lambda u^\gamma - \mathcal R_{\mu\nu}{}^{\alpha\beta}\right)\\
	&\quad+ \int d^3x \sqrt{g}F^{3/2} \left[
    c_2 \left(\frac{\mathcal R}{F} + \frac{(\partial_\mu F)^2}{2F^3}\right) + c_3 \left(\frac{{\mathcal R}_{\mu\nu}}{F} + \frac{3}{4} \frac{\partial_\mu F\partial_\nu F}{F^3}  - \frac{1}{2} \frac{\nabla_\mu \partial_\nu F}{F^2}\right)\frac{F^{\mu\alpha}  F^{\nu}{}_\alpha}{F^2}
    + c_4 \frac{(\nabla_\mu F^\mu{}_\alpha)^2}{F^3}
 \right] \\
	&\quad+ O(\partial^3)\, .
\end{split}
\end{equation}
\end{widetext}
Here we have defined $F \equiv \sqrt{\frac12 F_{\mu\nu}F^{\mu\nu}} = \sqrt{B^2 - E^2}$ and the unit vector $u^\mu \equiv \frac{1}{2F} \epsilon^{\mu\nu\lambda}F_{\nu\lambda}$. $\mathcal R_{\mu\nu}$ is the Ricci tensor of the manifold, and $\mathcal R = \mathcal R_\mu{}^\mu$. The first and third lines are present only for parity-violating CFTs. Given that we will expand around a constant magnetic field $F=B + \delta B(x)$, the appropriate derivative counting scheme is to take $F\sim \partial^0$. Each line is a given order in derivatives in this counting, from $O(\partial^{-1})$ to $O(\partial^2)$.  The construction of the generating functional above is technically similar to the construction of the effective action for conformal superfluids written in terms of a fluctuating gauge field~\citep{Cuomo2018}, the difference lies in the interpretation: in the present context the gauge field is not a fluctuating dynamical degree of freedom, but rather a background field. For a homogeneous magnetic field $F=B ={\rm const}$ and a flat metric $g_{\mu\nu}=\delta_{\mu\nu}$, one recovers Eq.~\eqref{eq:energy}.

This effective action controls the response functions of the CFT in a magnetic field, assuming the resulting system is gapped. The response will be universal, up to the theory-dependent Wilsonian coefficients $\nu,\,\chi_m,\,\zeta,\,\kappa,\,c_{2,3,4},\,$ etc.~which characterize the CFT. For example, $\chi_m$ denotes the magnetic susceptibility, and the $O(\partial^2)$ coefficients $c_{2,3,4}$ parametrize $q^2$ corrections to the susceptibility at finite wavevector $q$. The scale suppressing these corrections is the magnetic field (or magnetic length), i.e. observables can be computed in an expansion in $q^2/B$. When the theory is placed on a sphere, these $O(q^2/B)$ turn into $1/(BR^2) = 1/Q_B$ corrections to observables such as the free energy, see Eq.~\eqref{eq_DeltaQ_gapped}.

\subsection{Sphere free energy and defect monopoles}

Let us take $F$ to be proportional to the volume form on $S_2$: $F_{\theta\phi} = B \sin \theta$. First assuming invariance under reflection symmetry, the free-energy density is 
\begin{equation}
\mathcal F = -\frac{1}{V \beta} \log Z 
	       = c_1 B^{3/2} + \frac{1}{R^2}B^{1/2} \left(2 c_2 + c_3\right) + \cdots\, ,
\end{equation}
with $c_1\equiv \frac43 \chi_m$. The energy of the ground state on the sphere is $4\pi R^2 \mathcal F$, so that the dimension of the lightest operator of magnetic charge $Q_B = B R^2$ is
\begin{align}
\Delta &= 4\pi R^3 \mathcal F \notag \\
	&= 4\pi (BR^2 )^{3/2} + 4\pi \left(2 c_2 + c_3\right) (BR^2)^{1/2} + \cdots \label{eq_DeltaQ_gapped}\\
	&= 4\pi c_1 Q_B^{3/2} + 4\pi \left(2 c_2 + c_3\right) Q_B^{1/2} + O(Q_B^{-1/2})\notag
\end{align}
In particular, we note the absence of a $(Q_B)^0$ term. In the context of large (electric) charge $Q$ expansion, a $Q^0$ term arises from phonon fluctuations if the CFT enters a superfluid phase \cite{hellerman_on_2015}. Here we have instead assumed that the system is gapped, which leads to an expansion for $\Delta/Q_B^{3/2}$ in integer powers of $1/Q_B$. We shall verify the vanishing of the $Q_B^0$ term in specific CFTs in Section~\ref{sec:FreeEnergyCalcs}.

In CFTs that do not have reflection as a symmetry, one may expect $\Delta/Q_B^{3/2}$ to have an expansion in half-integer powers of $1/Q_B$ instead. One can check however that the contributions from both $\zeta$ and $\kappa$ terms above to the free energy vanish (this also holds in the presence of a holonomy $\int d\tau A_0 \in 2\pi \mathbb Z$, as long as $\kappa\in \mathbb Z$). We leave the study of higher-derivative parity-odd terms, and whether they contribute to the free energy, for future work.

\section{Dual Higgs mechanism}\label{sec_Higgs}

In the previous section, we assumed that a 3d CFT with a U(1) symmetry placed in a magnetic field enters a gapped phase, and derived consequences of this assumption. Here, we will argue that this scenario is fairly generic, at least for weakly coupled CFTs, using the action of $SL(2,\mathbb Z)$ on these theories \cite{Witten2003}.

Consider a CFT of interest, $\rm CFT_1$, which will be placed in a background magnetic field\footnote{For a theory with fermions on a general spacetime manifold, the background gauge field  should be taken to be a spin$_c$ connection. This will not a play a role in our present discussion.}. Its partition function can be written in terms of that of its $S$-dual partner, ${\rm CFT_2 } \equiv S({\rm CFT_1})$, as follows 
\begin{equation}\label{eq_Sdual}
Z_{\rm CFT_1}[A]
	= \int D a \, Z_{\rm CFT_2}[a] \, e^{\frac{i}{2\pi}\int adA}\, .
\end{equation}
This equation in fact defines the action of $S$ (or rather its inverse). In this formulation of the original {CFT$_1$}, the $U(1)$ current is carried by a dynamical gauge field $a$. Let us turn on a constant background magnetic field $\epsilon^{ij}\partial_i A_j$. This will source a chemical potential $\langle \int d\tau a_0\rangle \neq 0$ and relatedly a finite density for the current of $\rm CFT_2$. Assuming the CFT is weakly coupled, we can consider the dynamics of the `matter sector' in $\rm CFT_2$ and the dynamical gauge field $a$ separately. $\rm CFT_2$ is placed at finite density: we expect it to enter a superfluid phase. The weak coupling to the dynamical gauge field will however Higgs the full system, which is therefore gapped. 

Let us consider an example: the Wilson-Fisher CFT of $N$ complex bosons at large $N$. We couple the diagonal $U(1)$ current to a background gauge field:
\begin{equation}\label{eq_O2N}
\begin{split}
Z_{\rm CFT_1}[A] &= \int D\phi_i \, e^{i \int \mathcal L_1[\phi,A]}\, , \\
\mathcal L_1 &= \sum_{i=1}^N |D_{\! A}\phi_i|^2 - \lambda|\phi_i|^4\, .
\end{split}
\end{equation}
The $S$-dual is a bosonic gauge theory (QED)
\begin{equation}
\begin{split}
Z_{\rm CFT_2}[a] &= \int D\widehat\phi_i Db \, e^{i \int \mathcal L_2[\widehat\phi,b,a]}\, , \\
\mathcal L_2&=
     \sum_{i=1}^N |D_{ b}\widehat \phi_i|^2 - \lambda|\widehat\phi_i|^4 + \frac{1}{2\pi}bda\, , 
\end{split}
\end{equation}
which is known to enter a superfluid phase in a background chemical potential $a_0\neq 0$ \cite{sachdev_compressible_2012,delaFuente_large_2018}. Introducing $Z_{\rm CFT_2}$ in the path integral \eqref{eq_Sdual}, one expects to obtain instead a superconductor, which is gapped. In Sec.~\ref{sec:FreeEnergyCalcs}, we study the theory \eqref{eq_O2N} directly by computing the dimension of background monopole operators at large flux, and find that their scaling dimension indeed behaves as \eqref{eq_DeltaQ_gapped} (with no $Q_B^0$ piece), as expected for operators corresponding to gapped states. An analogous analysis holds for the Gross-Neveu-Yukawa fixed point with a large number of Dirac fermions; the large-flux expansion is again agreement with a gapped phase, see Sec.~\ref{sec:FreeEnergyCalcs}.

\section{Defect monopoles in free and large-$N$ CFTs}
\label{sec:FreeEnergyCalcs}
We now obtain the defect monopole scaling dimensions by studying the sphere free-energies in several QFTs of interest in a background magnetic field. First,
we study the case of non-interacting fermionic and bosonic theories,
and in the next subsection we consider interacting CFTs.

\subsection{Free CFTs}
Consider the free Dirac fermion CFT. The result, which also applies to the leading-order large-$N$ free energy
in the GN, QED, and QED-GN models~\citep{dupuis_2022}, is exactly given by 
\begin{equation}
F_{f}=-2\sum_{\ell=Q_{B}}^{\infty}\ell\sqrt{\ell^{2}-Q_{B}^{2}}.\label{eq:FreeEnergy_Fermions}
\end{equation}
Note that this expression corresponds to the unregularized free energy.
We work directly with the unregularized free energy, expand it for
large-$Q_{B}$, and then use Zeta-function regularization~\citep{Shiekh1990,NIST:DLMF,olver1997}
to regularize the ensuing sum. In the large-$Q_{B}$ limit, the free
energy then becomes 
\begin{align}
F_{f} & =-2\sum_{\ell=0}^{\infty}\left(\ell+Q_{B}\right)\sqrt{\left(\ell+Q_{B}\right)^{2}-Q_{B}^{2}}\nonumber \\
 & =-2^{\frac{3}{2}}\zeta\left(-1/2\right)Q_{B}^{3/2}-\frac{5}{\sqrt{2}}\zeta\left(-3/2\right)Q_{B}^{1/2}\nonumber \\
 & \quad-\frac{7\sqrt{2}}{16}\zeta\left(-5/2\right)Q_{B}^{-1/2}+O\left(Q_{B}^{-3/2}\right).\label{eq:FreeFermions_LargeQ}
\end{align}
The absence of a $Q_{B}^{0}$ term was discussed in Ref.~\cite{dupuis_2022},
where $NF_{f}$ is the leading-order result for the scaling dimension
of local quantum monopoles (not defects) in fermionic QED
where a U(1) gauge field couples to $N$ Dirac fermions. 
Furthermore, in the Dirac CFT, there are numerous degenerate defect
monopoles. Indeed, the magnetic flux through the sphere leads to $2|Q_{B}|N$
zero modes, where $N$ is the number of 2-component Dirac fermions.
For the defect monopole to be gauge-invariant (under background gauge
transformations), one must fill half of the zero modes. A large degeneracy
of distinct defect monopoles thus appears as $Q_{B}$ increases.

Now consider the free complex boson CFT. The (unregularized) free
energy corresponding to monopoles in this model~\citep{pufu_monopoles_2013}
is given by~\footnote{Note that, in Ref.~\citep{pufu_monopoles_2013} $Q_{B}$ is defined
to be an integer, and thus it corresponds to our $n$.}: 
\begin{equation}
F_{b}=\left.2\sum_{\ell=Q_{B}}^{\infty}\left(\ell+\frac{1}{2}\right)\sqrt{\left(\ell+\frac{1}{2}\right)^{2}+a_{Q_{B}}^{2}}\right|_{a_{Q_{B}}^{2}=-Q_{B}^{2}}.\label{eq:FreeEnergy_Bosons}
\end{equation}
In the large-$Q_{B}$ limit, Zeta regularization gives the following
form for the bosonic free-energy: 
\begin{align}
F_{b} & =2\sum_{\ell=0}^{\infty}\left(\ell+Q_{B}+\frac{1}{2}\right)\sqrt{\left(\ell+Q_{B}+\frac{1}{2}\right)^{2}-Q_B^2 }
\nonumber \\
 & =2^{\frac{3}{2}}\zeta\left(-1/2,1/2\right)Q_{B}^{3/2}+\frac{5}{\sqrt{2}}\zeta\left(-\frac{3}{2},\frac{1}{2}\right)Q_{B}^{1/2}\nonumber \\
 & \quad+\frac{7\sqrt{2}}{16}\zeta\left(-\frac{5}{2},\frac{1}{2}\right)Q_{B}^{-1/2}+O\left(Q_{B}^{-3/2}\right).\label{eq:FreeBosons_LargeQ}
\end{align}
Again, we find only half-integer powers of $Q_B$, in particular no constant term, in agreement with our general analysis for gapped states.

\subsection{Interacting theories}

In the limit of large but finite $N$, where $N$ denotes either the
number of fermions or bosons, the monopole scaling dimension can be
written as 
\begin{equation}
\Delta_{Q_{B}}=N\Delta_{Q_{B}}^{(0)}+\Delta_{Q_{B}}^{(1)}+\dotsb
\end{equation}
One can then consider the large-charge expansion of this expression.
Thus, in the limit $Q_{B}\rightarrow\infty$, we have 
\begin{align}
\Delta^{(0)} & =\alpha_{0}Q_{B}^{3/2}+\beta_{0}Q_{B}^{1/2}+\gamma_{0}+\dotsb\\
\Delta^{(1)} & =\alpha_{1}Q_{B}^{3/2}+\beta_{1}Q_{B}^{1/2}+\gamma_{1}+\dotsb
\end{align}
The parameter $\gamma$ thus has a large-$N$ expansion given by:~\footnote{Note, our nomenclature is slightly different than that of Ref.~\citep{delaFuente_large_2018}.}
\begin{equation}
\gamma=N\gamma_{0}+\gamma_{1}+\frac{1}{N}\gamma_{2}.
\end{equation}
The statement that $\gamma$ is a universal constant is tantamount
to having all the $\gamma_{i}$ coefficients vanish except for $\gamma_{1}$.
Since the leading-order free energies for QED-GN, QED, and
GN are equivalent to the exact free-fermion free energy given
in Eq.~\eqref{eq:FreeEnergy_Fermions}, the result in Eq.~\eqref{eq:FreeFermions_LargeQ}
shows that $\gamma_{0}=0$ for these interacting fermionic theories.
The same statements are equally valid for the bosonic theories considered
in this paper. Indeed, the $\mbox{CP}^{N-1}$ and $\mbox{O}(N)$ models have the
same leading-order free energies as the exact free-boson free energy
(with different expressions for $a_{Q_{B}}^{2}$). Hence, the result
in Eq.~\eqref{eq:FreeBosons_LargeQ} shows that $\gamma_{0}=0$ for
these interacting bosonic theories. 

Let us now turn to the computation of the defect dimensions, and the resulting
coefficients appearing in the expansion for 
two particular interacting CFTs, GN and O($N$). 
Ref.~\citep{dupuis_2022} obtained the large-$N$
monopole anomalous dimension for the fermionic models QED, QED-GN,
and QED-$Z_{2}$GN. The same analysis can be performed for the
case of GN by removing the dynamical U(1) gauge field. As for the O($N$) model,
the anomalous dimensions were obtained in Ref.~\citep{pufu_monopoles_2013}.
The anomalous dimensions obtained are shown in Table~\ref{tab:dimensions}
for a few small-$Q_{B}$ values, with an explanation of their computation
in Appendix~\ref{sec:scaling_dimensions} along with higher-$Q_{B}$.

Directly fitting the anomalous dimensions in Table~\ref{tab:Anomalous_Dimensions} yields
\begin{equation}
\begin{aligned}(\Delta_{Q_{B}}^{(1)})_{\rm GN} & =0.180158(3)Q_{B}^{3/2}+0.01009(5)Q_{B}^{1/2}\\
 & \quad{}+0.0001(1)-0.00033(6)Q_{B}^{-1/2},\\
(\Delta_{Q_{B}}^{(1)})_{\mbox{O}(N)} & =-0.140953(4)Q_{B}^{3/2}-0.0114(2)Q_{B}^{1/2}\\
 & \quad{} -.0003(4)+0.0015(4)Q_{B}^{-1/2}\\
  & \quad{} -0.0005(2)Q_{B}^{-1}.
\end{aligned}
\end{equation}
The number of terms in the large-$Q_{B}$ expansion is limited by
the number of anomalous dimensions and their precision. Adding more
terms to these fits makes the error bars on sub-leading coefficients
too large. The results show that $\gamma_1$ vanishes within the uncertainty bounds, and they also suggest $\gamma=0$ in both models.

\begin{figure}
\begin{centering}
\includegraphics[width=1\linewidth]{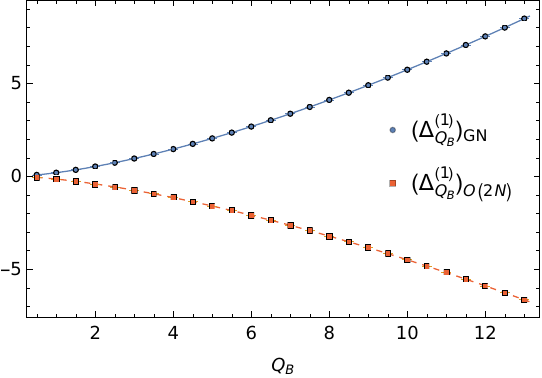} 
\par\end{centering}
\caption{Anomalous dimensions of defect monopoles in GN and O$(N)$ fitted
with the large-$Q_{B}$ expansion with powers $Q_{B}^{3/2}$ and $Q_{B}^{1/2}$
with coefficients shown in Eq.~\eqref{eq:lo_coefficients}. \label{fig:large_q_fit_1}}
\end{figure}

We employ a second fitting method to support the evidence that $\gamma_{1}=0$.
This method aims to alleviate the fact that we are working with a
data set with limited values of $Q_{B}$. We first fit $\Delta_{Q_{B}}^{(1)}$
with only the two leading terms, $\alpha_{1}Q_{B}^{3/2}+\beta_{1}Q_{B}^{1/2}$.
A set of coefficients are obtained by fitting many data sets where
smaller charges are progressively removed. A cubic fit is used to
model the tendency of the coefficients as the proportion of larger
charges gets higher, which gives 
\begin{equation}
\begin{cases}
\alpha_{1}=0.1801586(4),\quad\beta_{1}=0.010087(4),\; & \mbox{GN}\\
\alpha_{1}=-0.140957(1),\quad\beta_{1}=-0.01138(1),\; & \mbox{O}(N)
\end{cases}\label{eq:lo_coefficients}
\end{equation}
The results for the coefficients are within the error bars of the
coefficients obtained with the first method. The two leading-order
fitted terms nicely reproduce the data as shown in Fig.~\ref{fig:large_q_fit_1}.

\begin{figure}
\begin{centering}
\includegraphics[width=1\linewidth]{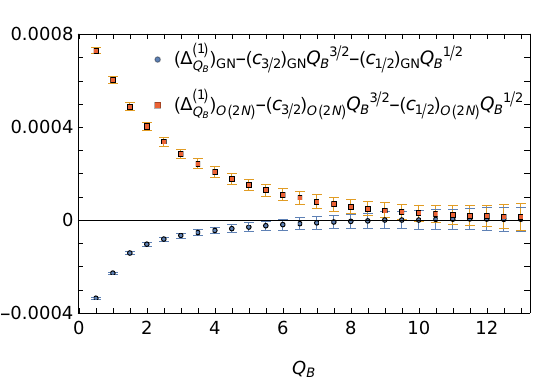} 
\par\end{centering}
\caption{Anomalous dimensions of monopoles in GN and O$(N)$ subtracted
by $Q_{B}^{3/2}$ and $Q_{B}^{1/2}$ fitted terms with coefficients
shown in Eq.~\eqref{eq:lo_coefficients}. \label{fig:large_q_fit_2}}
\end{figure}

Subtracting these two fitted terms from the anomalous dimensions,
$\Delta_{Q_{B}}^{(1)}-\alpha_{1}Q_{B}^{3/2}-\beta_{1}Q_{B}^{1/2}$,
we obtain the quantities shown in Fig.~\ref{fig:large_q_fit_2}.
As $Q_{B}$ gets larger, this quantity tends to $0$. Taking the value $\Delta_{Q_{B}}^{(1)}-\alpha_{1}Q_{B}^{3/2}-\beta_{1}Q_{B}^{1/2}$
for the largest charge, we obtain the following estimates for $\gamma$:
\begin{equation}
\gamma_{1}=\begin{cases}
0.00000(5), & \mbox{GN}\\
0.00001(7). & \mbox{O}(N)
\end{cases}
\end{equation}
These numerical results suggest yet again that the universal constant
in the large-$Q_{B}$ and large-$N$ expansions vanishes in both GN
and O$(N)$ models.
In Ref.~\cite{dupuis_2022}, another interacting theory of interest
was the QED-$Z_{2}$GN theory. As we did in going from QED-GN
to GN, we can consider QED-$Z_{2}$GN in the absence of a gauge
field, which results in $Z_{2}$GN. This theory has an interaction
term that favours fermion pairing given by $\phi^{*}\psi_{i}^{T}i\gamma_{2}\psi_{i}+\textrm{H.c.}$,
where $\phi$ is a complex scalar and the gamma matrix $\gamma_{2}$
is the Pauli matrix $\sigma_{y}$. The leading-order monopole scaling
dimension in QED-$Z_{2}$GN is the same as in QED. 
Furthermore, it was shown~\cite{dupuis_2022} that the anomalous scaling
dimension (the next-to-leading-order contribution) is given by $\Delta_{\textrm{QED-}Z_{2}\mbox{GN}}^{(1)}=\Delta_{\rm QED}^{(1)}+2\Delta_{\textrm{GN}}^{(1)}$.
Thus, in the absence of the dynamical gauge field, we find $\Delta_{Z_{2}\mbox{GN}}^{(1)}=2\Delta_{\mbox{GN}}^{(1)}$.
Hence, the large-$Q_{B}$ expansion for this theory (in the large-$N$
limit) can be simply obtained by multiplying the corresponding GN
results by a factor of two. As a result, the large $Q_{B}$-expansion
for the $Z_{2}$GN theory is given by 
\begin{equation}
\Delta_{Z_{2}\textrm{GN}}^{(1)}=2\alpha_{1}^{\mbox{GN}}Q_{B}^{3/2}+2\beta_{1}^{\mbox{GN}}Q_{B}^{1/2}+O\left(Q_{B}^{-1/2}\right).
\end{equation}
Here $\alpha_{1}^{\rm GN}$ and $\beta_{1}^{\rm GN}$ are the specific values
for $\alpha_{1}$ and $\beta_{1}$ in GN theory obtained above. Since
$\gamma_{1}$ was numerically observed (within uncertainty bounds)
to vanish in GN, this implies that it also vanishes in the large-$Q_{B}$
expansion of $Z_{2}$GN. Thus, $Z_{2}$GN represents another interacting
theory where $\gamma=0$ is expected to hold.

\subsection{Chiral Heisenberg Gross-Neveu CFT}
Another GN CFT that received considerable attention in the study
of quantum phase transitions is the chiral Heisenberg Gross-Neveu
(cHGN) model~\cite{meng_monte_2019,zerf_critical_2019,dupuis_transition_2019},
$\delta\sim-\vec{\sigma}\cdot\bar{\Psi}\vec{\phi}\Psi$. The interaction
is magnetic-spin dependent with the Pauli matrices $\vec{\sigma}$
acting on the magnetic spin (flavour) subspace. In GN and QED-GN models
where the fermionic interaction is symmetric, monopoles are described
at leading order in $1/N$ by the same state as in the free-fermion
CFT. This is not the case for magnetically charged monopoles in $\mbox{cHGN}$
and QED-cHGN where the auxiliary field has a non-vanishing expectation
value. Even with this difference, it still remains that ${\gamma=0}$
at leading order in the cHGN model, leaving open the possibility
that $\gamma$ may also vanish in this model. Obtaining anomalous
scaling dimensions of cHGN monopoles would allow testing this
possibility. Given the important role that magnetic spin plays by
breaking the degeneracy of monopoles \cite{dupuis_monopole_2021},
inspecting how $\gamma$ is affected by monopole magnetic spin in
both cHGN and QED-cHGN models could add an interesting and
enlightening twist in the study of large-charge expansions.

\subsection{Convexity of defect dimensions}

It was conjectured~\citep{aharony_convexity_2021} that the scaling
dimensions of the lightest operators charged under a global symmetry
obey convexity. While counterexamples have been found~\citep{Sharon2023},
convexity is often observed and may serve as a useful reference behavior.
Generalizing this observation, we expect that convexity will hold
for the defect scaling dimensions in a large class of theories: 
\begin{equation}
\Delta^{{\rm defect}}((n_{1}+n_{2})n_{0})\geq\Delta^{{\rm defect}}(n_{1}n_{0})+\Delta^{{\rm defect}}(n_{2}n_{0}),
\end{equation}
for a positive integer $n_{0}$ of order 1. Here, $n_{0},n_{1},n_{2}$
are integers, and in our notation $\Delta^{\rm defect}(2Q_{B})\equiv\Delta_{Q_{B}}$.
Convexity is justified for holographic theories \cite{aharony_convexity_2021}
by the expectation that gravity must be the weakest force in UV complete
descriptions of quantum gravity, so that there should exist self-repulsive
(magnetically) charged matter \cite{Arkani-Hamed2007}. We find
that convexity is obeyed for defect monopole operators in O($N$) and
GN models.

\begin{table}[H]
\centering{}\caption{Anomalous dimensions of defect monopole operators at next-to-leading
in $1/N$ in the GN and O($N$) models. The scaling dimension is
determined via $N\Delta_{Q_{B}}^{\left(0\right)}+\Delta_{Q_{B}}^{\left(1\right)}$.
\label{tab:dimensions}}
\begin{ruledtabular} %
\begin{tabular}{ccc}
$Q_{B}$  & $\Delta_{Q_{B},\mbox{GN}}^{\left(1\right)}$  & $\Delta_{Q_{B},\mbox{O}(N)}^{\left(1\right)}$ \vspace{0.25em}
 \tabularnewline
\hline 
$1/2$  & $0.0704917(8)$  & $-0.0571528$ \tabularnewline
$1$  & $0.190017(2)$  & $-0.1517343$ \tabularnewline
$3/2$  & $0.343185(4)$  & $-0.2724039$ \tabularnewline
$2$  & $0.523727(5)$  & $-0.4143773$ \tabularnewline
$5/2$  & $0.728007(7)$  & $-0.5748376$ \tabularnewline
$3$  & $0.953537(8)$  & $-0.7518602$ \tabularnewline
\end{tabular}\end{ruledtabular} 
\end{table}

\section{Alternatives: Metals and non-Fermi Liquids} \label{sec:nfl}

We have argued that many CFTs become gapped when placed in a magnetic field -- we expect this scenario to be fairly generic, in analogy with the large charge proposal of \cite{hellerman_on_2015} that many CFTs become superfluids when placed in a chemical potential. However, similar to the superfluid proposal, there are alternatives. In this section, we discuss several examples of CFTs that do not become gapped when placed in a magnetic field.

\subsection{A Dirac fermion coupled to Chern-Simons}
We consider a (2-component) Dirac fermion coupled to a fluctuating gauge field with Chern-Simons level $U(1)_{-k+\frac12}$
\begin{equation}
S = \int d^3 x \, \bar\psi i \slashed D_a \psi - \frac{k}{4\pi} \int ada + \frac{1}{2\pi} \int C d a\, .
\end{equation}
This is QED$_3$ with a single Dirac fermion and CS coupling for the photon. 
$C_\mu$ is a background gauge field, coupled to the $U(1)$ current (magnetic flux) of the theory. The operators charged under this global U(1) symmetry are monopoles or instantons of $a_\mu$.
When $k\gg 1$, the theory becomes weakly coupled. This model was studied in the large charge context in Ref.~\cite{Cuomo2021}. It can be shown to become a superfluid when a finite density is generated by turning on a chemical potential for the conserved charge, at least when $k\gg 1$. It is a relativistic cousin of the well-known `anyon superfluid' \cite{Chen:1989xs}.

We would now like to study what happens to this theory in a background magnetic field 
\begin{equation}
\partial_{i} C_{j} - \partial_j C_i = B \epsilon_{ij}\, .
\end{equation}
Since we are interested in the behavior of CFT operators with large magnetic charge but vanishing $U(1)$ charge, we also set the average density for the global $U(1)$ symmetry to zero
\begin{equation}
\int d^2x\, \left(\partial_i a_j - \partial_j a_i \right)
	=  \int d^2x\,  b \epsilon_{ij}= 0\, .
\end{equation}
The effective magnetic field $b$ that the fermions feel therefore vanishes. This is the crucial property that leads to a gapless phase at finite external magnetic field. 
The constrained variable $a_0$ has an equation of motion that implements flux attachment
\begin{equation}
0 = \bar\psi \gamma^0 \psi + \frac{1}{2\pi} (kb - B)\, .
\end{equation}
We therefore have a `finite density' of fermions $j_\psi^0 \equiv \bar\psi \gamma^0 \psi = \frac{1}{2\pi} B$ coupled to a dynamical gauge field. The radius of the circular Fermi surface is
\begin{align}
    p_F = \sqrt{2|B|}
\end{align}
Sending first $k\to \infty$, the Fermi gas becomes decoupled from the gauge field, and the system is gapless. This already offers an alternative to the more common scenario considered earlier that CFTs placed in a magnetic field become gapped.

A more interesting alternative arises by turning on a small coupling $1/k$: the model is then very similar to the Halperin-Lee-Read \cite{Halperin1992} (or Nayak-Wilczek \cite{Nayak1994}) model of a non-Fermi liquid (NFL), which is expected to be stable against superconductivity. Working in Coulomb gauge $\nabla_i a_i = 0$, one finds at one-loop that $a_0$ gets Debye-screened by the Fermi surface, and $a_i$ is Landau damped 
\begin{equation}
\langle aa\rangle (\omega,q)
	\sim \frac{1}{q^2 + \gamma \frac{|\omega|}{|q|}}\, .
\end{equation}
Interaction corrections to this are then {\em relevant}, and one expects a fully fledged NFL. Perturbatively, the superconducting instability seems to be absent due to the fact that $a_i$ couples with opposite signs to two opposing patches. We note that this CFT evades the argument of Sec.~\ref{sec_Higgs} because its S-dual CFT, a free Dirac fermion, does not become a superfluid at finite density (instead, it becomes a Fermi gas).

\subsubsection{Background monopoles at large flux}
We now consider the large flux limit of the background monopoles when $k\to\infty$. Working on the sphere, 
one obtains a finite density of non-interacting fermions proportional to $B$. This state maps to a large magnetic charge operator, with dimension 
\begin{equation}\label{eq_Fermigas}
\Delta = \# Q_B^{3/2} + \# Q_{B}^{1/2} + O(Q_B^{-1/2})\, .
\end{equation}
The absence of an $O(Q^0)$ term for the free Fermi gas was found in \cite{komargodski_spontaneously_2021} by carefully constructing the lightest large-charge operators. It actually simply follows from the stronger statement that the density two-point function of a free Fermi gas $\langle j^0 j^0\rangle(\omega=0,q)$ is {\em analytic} near $q=0$.
Indeed, as discussed in Sec.~\ref{sec:functional}, finite-wavevector corrections to response functions translate into $1/Q$ corrections to $\Delta$ (the units are made up with the density for large charge, and magnetic field for large magnetic charge). This argument further forbids any power in \eqref{eq_Fermigas} that is not half-integer, including logarithms. Analyticity is however an artefact of the free Fermi gas; a Fermi liquid should have non-analytic corrections $\langle j^0 j^0\rangle(\omega=0,q)\sim 1 + |q|^\alpha + \cdots$. These corrections are in fact subtle, due to partial cancellations in fermion loops: while the naively expected correction with $\alpha=d-1$ is absent \cite{PhysRevB.68.155113}, scaling arguments that account for these cancellations suggest that the leading correction should have $\alpha = d+1$ (as in superfluids) \cite{Delacretaz:2022ocm}.
This implies that Fermi liquids should have a constant $O(Q^0)$ correction to the dimension $\Delta$ (with an additional $\log Q$ in odd spatial dimensions), like for superfluids.

When the level $k$ is instead large but finite, the finite-density dynamics becomes strongly coupled, as discussed above. Because we do expect non-analyticities in the static correlator $\langle j^0 j^0\rangle(\omega=0,q)$, the behavior of the scaling dimension should be more subtle than in Eq.~\eqref{eq_Fermigas}; however, we can only speculate on their form because $\langle j^0 j^0\rangle(\omega=0,q)$ is not currently known for non-Fermi liquids.

\subsection{Holographic CFTs}

Holographic CFTs at $N=\infty$ with a $U(1)$ symmetry already offer a counter-example to the large charge proposal of \cite{hellerman_on_2015}: the typical zero temperature finite density state is an extremal Reissner-Nordstrom black hole rather than a superfluid. Extremal black holes are distinguished from conventional finite density phases of matter by a number of properties, including their finite zero temperature entropy. 

When placed in a background magnetic field, holographic CFTs typically form extremal {\em magnetic} black holes, which also feature a finite zero-temperature entropy \cite{Hartnoll2007}. These solutions are simply the electric-magnetic duals $F_{\mu\nu} \to \epsilon_{\mu\nu\lambda\sigma} F^{\lambda\sigma}$ of usual charged black holes (in fact, a CFT whose bulk is self-dual under electric-magnetic duality must have an identical spectrum of large charge and large magnetic charge operators). These therefore provide a dramatic alternative to the gapped phase that many CFTs realize in a magnetic field. These states, as well as their generalization to extremal `dyonic' black holes carrying both magnetic and electric charge, have a near-horizon description in terms of Jackiw-Teitelboim gravity~\cite{Moitra2019}, an effective two-dimensional theory of gravity that also enters in the low-energy description of the SYK model. Quantum $1/N$ corrections to their equation of state have been subject of recent interest (see, e.g., Ref.~\cite{Heydeman2022}).

\section{Conclusion}
\label{sec:Conclusion}

Several open questions remain. First and foremost, can one classify more completely the possible phases obtained by applying a magnetic field to CFTs? In particular, the stability of the putative non-Fermi liquid that arose from the gauged Dirac fermion should be investigated as a function of the Chern-Simons coupling. An analogous analysis is needed for the extremal black holes generated in holographic CFTs. 

An even larger set of possibilities arises by allowing for a chemical potential $\mu$ in combination with a magnetic field $B$, as the CFT equation of state can nontrivially depend on the dimensionless combination $\mu^2/B$, or equivalently the filling fraction $\nu = Q/Q_B$. We expect many phases may be realized as a function of filling for a given CFT, such as fractional quantum Hall or Wigner crystal states. This would offer a new interesting connection between quantum Hall and 2+1d CFTs, beyond the well known ones with 1+1d CFTs \cite{hansson_QH_2017}. In the context of the large charge expansion of CFTs, ratios of large quantum numbers have already lead to interesting possibilities along those lines \cite{Cuomo2018, Delacretaz:2020nit, Cuomo2023}.

\begin{acknowledgements}
We thank Shai Chester for helpful discussions. 
{W.W.-K.} and {R.B.} were funded by a Discovery Grant from NSERC, a Canada Research Chair, a grant from the Fondation Courtois, and a ``\'Etablissement de nouveaux chercheurs et de nouvelles chercheuses universitaires'' grant from FRQNT.
W.W.-K.\/ is supported by a Chair of the Institut Courtois.
 {R.B.} also thanks Dartmouth College for support. {\'E.D.} was funded by an Alexander Graham Bell CGS from NSERC. 
\end{acknowledgements}

\onecolumngrid
\appendix
\label{sec:Appendices}

\section{Generating functional for gapped phases}\label{app_gapped}
\subsection{Formulation}

In this appendix, we construct the generating functional $\log Z[A_\mu, g_{\mu\nu}]$ for CFTs that become gapped in a background magnetic field. Because the phase is gapped, the partition function is a local functional of background fields $A_\mu,\,g_{\mu\nu}$. It must be invariant under gauge transformations and diffeomorphisms. We will also impose invariance under Weyl transformations
\begin{equation}
\begin{split}
g_{\mu\nu}(x)&\to \Omega(x)^2g_{\mu\nu}(x),\\
A_{\mu}(x)&\to A_{\mu}(x)\, .
\end{split}
\end{equation}
Note that $|F|\equiv (\frac12 F_{\mu\nu}F_{\alpha\beta}g^{\mu\alpha}g^{\nu\beta})^{1/2}$ has weight -2 under Weyl transfromations $|F|\to |F| \Omega^{-2}$; one can therefore efficiently construct Weyl invariant terms by using the Weyl invariant metric \cite{cuomo2020large}
\begin{equation}
\hat g_{\mu\nu}\equiv g_{\mu\nu} |F|\, ,
\end{equation}
to write down diffeomorphism invariant terms. The only such term at zeroth order in derivatives is
\begin{equation}
c_1 \sqrt{\hat g}
	= c_1 \sqrt{g} |F|^{3/2}\, .
\end{equation}
For parity-invariant theories, the next terms enter at second order in derivatives and read:
\begin{subequations}
\begin{align}
c_2 \sqrt{\hat g} \hat {\mathcal R}
	&= c_2 \sqrt{g}|F|^{3/2} \left(\frac{\mathcal R}{|F|} + \frac{1}{2|F|^3}(\partial_\mu|F|)^2\right) \\
c_3 \sqrt{\hat g} \hat {\mathcal R}_{\mu\nu}\hat F^{\mu\alpha} \hat F^{\nu}{}_\alpha
	&=  c_3 \sqrt{g}|F|^{3/2} \left[\frac{{\mathcal R}_{\mu\nu}}{|F|} + \frac{3}{4} \frac{\partial_\mu|F|\partial_\nu|F|}{|F|^3}  - \frac{1}{2} \frac{\nabla_\mu \partial_\nu |F|}{|F|^2}\right]\frac{F^{\mu\alpha}  F^{\nu}{}_\alpha}{|F|^2}\\
c_4 \sqrt{\hat g} (\hat \nabla_\mu \hat F^{\mu \alpha})^2 
	&= c_4 \sqrt{g} |F|^{3/2} \frac{(\nabla_\mu F^\mu{}_\alpha)^2}{|F|^3}
\end{align}
\end{subequations}
The Ricci tensor of a Weyl-transformed metric can be found, e.g., in \cite{nakahara2003geometry}. We have used the Bianchi identity to remove one possible term $(\hat \nabla_{[\mu}F_{\nu\lambda]})^2$. The additional term $c_4$ compared to the large charge EFT \cite{hellerman_on_2015} arises because $F_{\mu\nu}$ is a background field, which need not satisfy an equation of motion.

\subsection{CFTs without parity}
When parity is not a symmetry, the generating functional can also contain a Chern-Simons term, which counts as $-1$ derivative in our counting scheme:
\begin{equation}
    \frac{\nu}{4\pi} \epsilon^{\mu\nu\lambda}A_\mu \partial_\nu A_\lambda\, ,
\end{equation}
At the one derivative level, there are additionally two terms that are also gauge, diffeomorphism, and Weyl-invariant \cite{Golkar2014,Golkar2015} :
\begin{equation}
    \zeta \, |F| \epsilon^{\mu\nu\lambda} u_\mu \partial_\nu u_\lambda\, , \qquad
	\frac{\kappa}{8\pi}\,\epsilon^{\mu\nu\lambda}\epsilon_{\alpha\beta\gamma} A_\mu u^\alpha \left(\nabla_\nu u^\beta \nabla_\lambda u^\gamma - \mathcal R_{\mu\nu}{}^{\alpha\beta}\right)\, ,
\end{equation}
where we defined $u^\mu \equiv \frac{1}{2|F|} \epsilon^{\mu\nu\lambda}F_{\nu\lambda}$ to simplify notation. The second term is only gauge and Weyl invariant up to a total derivative, and has interesting physical properties. It is captures the `shift' of relativistic quantum Hall phases~\citep{Golkar2014}, and is proportional to the Hall viscosity of the CFT in a magnetic field. This term was studied in the large charge context of relativistic superfluid EFTs in~\citep{Cuomo2021}.

Collecting all terms so far, one finds the following generating functional for gapped phases of 2+1d CFTs with a $U(1)$ symmetry: 
\begin{equation}
\label{eq:logZ_App}
\begin{split}
\log Z[A,g]
	&= \frac{\nu}{4\pi} \int d^3 x \, \epsilon^{\mu\nu\lambda} A_\mu \partial_\nu A_\lambda \\
	&+ \frac43 \chi_M \int d^3 x \sqrt{g} F^{3/2}\\
	&+ \zeta \int d^3 x \, F \epsilon^{\mu\nu\lambda} u_\mu \partial_\nu u_\lambda
	+ \frac{\kappa}{8\pi} \int d^3 x \,\epsilon^{\mu\nu\lambda}\epsilon_{\alpha\beta\gamma} A_\mu u^\alpha \left(\nabla_\nu u^\beta \nabla_\lambda u^\gamma - \mathcal R_{\mu\nu}{}^{\alpha\beta}\right)\\
	&+ \int d^3x \sqrt{g}F^{3/2} \left[
    c_2 \left(\frac{\mathcal R}{F} + \frac{(\partial_\mu F)^2}{2F^3}\right) + c_3 \left(\frac{{\mathcal R}_{\mu\nu}}{F} + \frac{3}{4} \frac{\partial_\mu F\partial_\nu F}{F^3}  - \frac{1}{2} \frac{\nabla_\mu \partial_\nu F}{F^2}\right)\frac{F^{\mu\alpha}  F^{\nu}{}_\alpha}{F^2}
    + c_4 \frac{(\nabla_\mu F^\mu{}_\alpha)^2}{F^3}
 \right] \\
	&+ O(\partial^3)
\end{split}
\end{equation}
Each line in the expansion above is at a given order in the derivative expansion, from $O(\partial^{-1})$ to $O(\partial^2)$. Parity-invariant CFTs will have $\nu,\,\zeta,\, \kappa = 0$.

\section{Monopole scaling dimension in specific models}

\subsection{Formulation}

The Euclidean action for GN is given by

\begin{equation}
S_{\rm GN}\left[\mathcal{A}^{Q_{B}}\right]=\int d^{3}r\sqrt{g}\left[-\bar{\Psi}\left(\slashed{D}_{\mathcal{A}^{Q_{B}}}+\phi\right)\Psi\right]+\dots.
\end{equation}
Here, $\Psi=\left(\psi_{1},\psi_{2},\dots\psi_{N}\right)^{\intercal},$
where each flavor is a two-component Dirac fermion, and the covariant
derivative with the external gauge field acting on the fermions is
given by 
\begin{equation}
\slashed{D}_{\mathcal{A}^{Q_{B}}}=\gamma^{\mu}\left(\nabla_{\mu}-i\mathcal{A}_{\mu}^{Q_{B}}\right).
\end{equation}
A symmetric fermion mass $\langle\bar{\Psi}\Psi\rangle\neq0$ is condensed
for a sufficiently strong coupling. The effective action at the quantum
critical point (QCP) is 
\begin{equation}
S_{\text{eff}}^{\rm GN}=-N\ln\det\left(\slashed{D}+\phi\right).
\end{equation}
The case of free fermions is obtained when we remove the GN interaction,
that is, by removing the contribution of the auxiliary field $\phi$.
This allows one to obtain the effective action used in the state-operator
correspondence to study the defect monopoles 
\begin{equation}
S_{\text{eff}}\left[\mathcal{A}^{Q_{B}}\right]=-N\times\begin{cases}
\log\det\left(\slashed{D}_{\mathcal{A}^{Q_{B}}}+\phi\right) & \mbox{GN}\\
\log\det\left(\slashed{D}_{\mathcal{A}^{Q_{B}}}\right) & \text{Free fermions}
\end{cases}.
\end{equation}

The Euclidean action for the $\mbox{O}(N)$ model is given by 
\begin{align}
S_{\mbox{O}(N)}\left[\mathcal{A}^{Q_{B}}\right] & =\int d^{3}r\sqrt{g}\left[\left|D_{\mathcal{A}^{Q_{B}}}z\right|^{2}+\frac{\mathcal{R}}{8}|z|^{2}+i\lambda\left(\left|z\right|^{2}-\frac{2N}{2\mathfrak{g}}\right)\right],
\end{align}
where $z=\left(z_{1},z_{2},\dots z_{N}\right)^{\intercal}$ and each
flavor is a complex scalar boson, $R$ is the Ricci scalar (with $\mathcal{R}=0$
and $\mathcal{R}=2$ on $\mathbb{R}^{3}$ and $\mathbb{R}\times S^{2}$
respectively), and the covariant derivative (with a background gauge
field) acting on the bosons is given by 
\begin{equation}
D_{\mathcal{A}^{Q_{B}}}=\nabla_{\mu}-i\mathcal{A}_{\mu}^{Q_{B}}.
\end{equation}
At the quantum critical point (defined on the flat spacetime), the
effective action is given by 
\begin{equation}
S_{\text{eff}}^{\mbox{O}(N)}=N\ln\det\left\{ -D^{2}+i\lambda\right\} .
\end{equation}
On the $\mathbb{R}\times S^{2}$ with the contribution of the Ricci
scalar, this becomes 
\begin{equation}
S_{\text{eff}}\left[\mathcal{A}^{Q_{B}}\right]=N\times\begin{cases}
\ln\det\left(-\left|D_{\mathcal{A}^{Q_{B}}}\right|^{2}+\frac{1}{4}+i\lambda\right), & \mbox{O}(N).\\
\ln\det\left(-\left|D_{\mathcal{A}^{Q_{B}}}\right|^{2}+\frac{1}{4}\right), & \text{free bosons},
\end{cases}
\end{equation}
where the self-interaction is removed in the case of the free boson.

\subsection{Monopole scaling dimensions \label{sec:scaling_dimensions}}

The monopole scaling dimensions in the O($N$) and GN models can
be deduced from the results in the literature. The O($N$) monopole
scaling dimension was explicitly obtained in Ref.~\citep{pufu_monopoles_2013}.
In the case of the GN model, anomalous dimensions of monopoles
were obtained for the QED-GN model~\citep{dupuis_2022}, and we
simply need to remove gauge fluctuations. In this appendix, we give
a brief overview of the results needed to obtain the points in Fig.\ref{fig:large_q_fit_1}.

In both cases, the anomalous dimension involves only the contribution
from the kernel of the auxiliary field decoupling the interaction.
Thus, the anomalous dimension can be written generically as 
\begin{equation}
\Delta_{Q_{B}}^{\left(1\right)}=\frac{1}{2}\int_{-\infty}^{\infty}\frac{d\omega}{2\pi}\sum_{\ell=0}^{\infty}\left(2\ell+1\right)\ln\left[\dfrac{D_{\ell}^{Q_{B}}\left(\omega\right)}{D_{\ell}^{0}\left(\omega\right)}\right].\label{eq:anomalous_dimension}
\end{equation}
The momentum-space coefficient is obtained by projecting the real-space
kernel on spherical harmonics 
\begin{align}
D_{\ell}^{Q_{B}}\left(\omega\right) & =\frac{4\pi}{2\ell+1}\int_{r}e^{i\omega\tau}D^{Q_{B}}\left(r,0\right)\sum_{m=-\ell}^{\ell}Y_{\ell m}^{*}\left(\hat{n}\right)Y_{\ell m}\left(\hat{z}\right)=\int_{r}e^{i\omega\tau}D^{Q_{B}}\left(r,0\right)P_{\ell}\left(x\right),
\end{align}
where we used the addition theorem. The real-space kernel depends
on the specific model: 
\begin{equation}
D^{Q_{B}}\left(r,r^{\prime}\right)=\begin{cases}
-\text{tr}\left[G_{Q_{B}}^{f}\left(r,r^{\prime}\right)G_{Q_{B}}^{f\dagger}\left(r,r^{\prime}\right)\right] & \mbox{GN}\\
G_{Q_{B}}^{b}(r,r')G_{Q_{B}}^{b*}(r,r') & \mbox{O}(N)
\end{cases}.
\end{equation}
Here, the Green's functions obey the following equations of motion:
\begin{align}
i\slashed{D}_{\mathcal{A}^{Q_{B}}}^{S^{2}\times\mathbb{R}}\left(r\right)G_{Q_{B}}^{f}\left(r,r'\right) & =-\delta\left(r-r^{\prime}\right)\\
\left(\Bigl|D_{\mu}^{\left(\mathcal{A}^{Q_{B}}\right)}(r)\Bigr|^{2}-\left(a_{Q_{B}}^{2}+Q_{B}^{2}+\frac{1}{4}\right)\right)G_{Q_{B}}^{b}(r,r') & =-\delta(r-r^{\prime}),
\end{align}
where we reformulated the O$(N)$ saddle point parameter as $\langle\lambda\rangle=a_{Q_{B}}^{2}+Q_{B}^{2}$
and used that $\langle\phi\rangle=0$ in the case of GN. The value
of $a_{Q_{B}}^{2}$ is determined by the saddle-point equation 
\begin{equation}
\sum_{\ell=Q_{B}}\left(\frac{\ell+1/2}{\sqrt{(\ell+1/2)^{2}+a_{Q_{B}}^{2}}}-1\right)=Q_{B}.
\end{equation}
The saddle-point equation is solved for values of the topological
charge up to $Q_{B}=13$.

\begin{table}[H]
\caption{Saddle-point results in GN and O$(N)$ with multiples values of
$Q_{B}$.}
\begin{minipage}[t]{0.48\columnwidth}%
\begin{ruledtabular} %
\begin{tabular}{cccc}
$Q_{B}$  & $\Delta_{Q_{B},\mbox{GN}}^{\left(0\right)}$  & $\Delta_{Q_{B},\mbox{O}(N)}^{\left(0\right)}$ \vspace{0.25em}
  & $a_{Q_{B}}^{2}$\tabularnewline
\hline 
$1/2$  & 0.265096  & 0.124592  & -0.449806\tabularnewline
$1$  & 0.673153  & 0.311095  & -1.397830\tabularnewline
$3/2$  & 1.186434  & 0.544069  & -2.845457\tabularnewline
$2$  & 1.786901  & 0.815788  & -4.792936\tabularnewline
$5/2$  & 2.463451  & 1.121417  & -7.240344\tabularnewline
$3$  & 3.208372  & 1.457570  & -10.187714\tabularnewline
$7/2$  & 4.015906  & 1.821708  & -13.635059\tabularnewline
$4$  & 4.881539  & 2.211833  & -17.582390\tabularnewline
$9/2$  & 5.801615  & 2.626323  & -22.029709\tabularnewline
$5$  & 6.773088  & 3.063825  & -26.977021\tabularnewline
$11/2$  & 7.793375  & 3.523189  & -32.424327\tabularnewline
$6$  & 8.860246  & 4.003422  & -38.371628\tabularnewline
$13/2$  & 9.971750  & 4.503655  & -44.818926\tabularnewline
\end{tabular}\end{ruledtabular} %
\end{minipage}$\hfill$%
\begin{minipage}[t]{0.48\columnwidth}%
\begin{ruledtabular} %
\begin{tabular}{cccc}
$Q_{B}$  & $\Delta_{Q_{B},\mbox{GN}}^{\left(0\right)}$  & $\Delta_{Q_{B},\mbox{O}(N)}^{\left(0\right)}$ \vspace{0.25em}
  & $a_{Q_{B}}^{2}$\tabularnewline
\hline 
$7$  & 11.126163  & 5.023119  & -51.766222\tabularnewline
$15/2$  & 12.321948  & 5.561128  & -59.213515\tabularnewline
$8$  & 13.557721  & 6.117064  & -67.160806\tabularnewline
$17/2$  & 14.832227  & 6.690367  & -75.608096\tabularnewline
$9$  & 16.144324  & 7.280526  & -84.555385\tabularnewline
$19/2$  & 17.492965  & 7.887074  & -94.002672\tabularnewline
$10$  & 18.877186  & 8.509578  & -103.949959\tabularnewline
$21/2$  & 20.296094  & 9.147641  & -114.397245\tabularnewline
$11$  & 21.748862  & 9.800891  & -125.344530\tabularnewline
$23/2$  & 23.234719  & 10.468985  & -136.791815\tabularnewline
$12$  & 24.752943  & 11.151598  & -148.739099\tabularnewline
$25/2$  & 26.302859  & 11.848430  & -161.186382\tabularnewline
$13$  & 27.883833  & 12.559195  & -174.133665\tabularnewline
\end{tabular}\end{ruledtabular} %
\end{minipage}
\end{table}

\begin{table}[H]
\caption{Anomalous dimensions of defect monopole operators in large-$N$ for
the GN and O$(N)$ models. The scaling dimension is determined
via $N\Delta_{Q_{B}}^{\left(0\right)}+\Delta_{Q_{B}}^{\left(1\right)}$.
\label{tab:Anomalous_Dimensions}}

\begin{minipage}[t]{0.48\columnwidth}%
\begin{ruledtabular} %
\begin{tabular}{ccc}
$Q_{B}$  & $\Delta_{Q_{B},\mbox{GN}}^{\left(1\right)}$  & $\Delta_{Q_{B},\mbox{O}(N)}^{\left(1\right)}$ \vspace{0.25em}
 \tabularnewline
\hline 
$1/2$  & $0.0704917(8)$  & $-0.0571528$ \tabularnewline
$1$  & $0.190017(2)$  & $-0.1517343$ \tabularnewline
$3/2$  & $0.343185(4)$  & $-0.2724039$ \tabularnewline
$2$  & $0.523727(5)$  & $-0.4143773$ \tabularnewline
$5/2$  & $0.728007(7)$  & $-0.5748376$ \tabularnewline
$3$  & $0.953537(8)$  & $-0.7518602$ \tabularnewline
$7/2$  & $1.19850(1)$  & $-0.9440208$ \tabularnewline
$4$  & $1.46140(2)$  & $-1.1502101$ \tabularnewline
$9/2$  & $1.74115(2)$  & $-1.3695321$ \tabularnewline
$5$  & $2.03676(2)$  & $-1.6012435(4)$ \tabularnewline
$11/2$  & $2.34743(2)$  & $-1.8447143(7)$ \tabularnewline
$6$  & $2.67247(2)$  & $-2.0994022(11)$ \tabularnewline
$13/2$  & $3.01126(3)$  & $-2.3648334(16)$ \tabularnewline
\end{tabular}\end{ruledtabular} %
\end{minipage}$\hfill$%
\begin{minipage}[t]{0.48\columnwidth}%
\begin{ruledtabular} %
\begin{tabular}{ccc}
$Q_{B}$  & $\Delta_{Q_{B},\mbox{GN}}^{\left(1\right)}$  & $\Delta_{Q_{B},\mbox{O}(N)}^{\left(1\right)}$ \vspace{0.25em}
 \tabularnewline
\hline 
$7$  & $3.36326(3)$  & $-2.6405897(22)$ \tabularnewline
$15/2$  & $3.72800(3)$  & $-2.9262981(31)$ \tabularnewline
$8$  & $4.10505(3)$  & $-3.221624(4)$ \tabularnewline
$17/2$  & $4.49402(4)$  & $-3.526262(6)$ \tabularnewline
$9$  & $4.89454(4)$  & $-3.839939(7)$ \tabularnewline
$19/2$  & $5.30631(4)$  & $-4.162399(10)$ \tabularnewline
$10$  & $5.72902(4)$  & $-4.493410(12)$ \tabularnewline
$21/2$  & $6.16239(4)$  & $-4.832756(15)$ \tabularnewline
$11$  & $6.60616(4)$  & $-5.180239(19)$ \tabularnewline
$23/2$  & $7.06011(5)$  & $-5.535671(24)$ \tabularnewline
$12$  & $7.52400(5)$  & $-5.898878(29)$ \tabularnewline
$25/2$  & $7.99763(5)$  & $-6.269699(35)$ \tabularnewline
$13$  & $8.48080(5)$  & $-6.64798(4)$ \tabularnewline
\end{tabular}\end{ruledtabular}%
\end{minipage}
\end{table}

\subsubsection{Zero topological charge}

With a vanishing topological charge, the Green's functions in both
models have a closed form 
\begin{align}
G_{0}^{f}(r,r') & =\frac{i}{4\pi}\frac{\vec{\gamma}\cdot\left(e^{\frac{1}{2}\left(\tau-\tau^{\prime}\right)}\hat{n}-e^{-\frac{1}{2}\left(\tau-\tau^{\prime}\right)}\hat{n}^{\prime}\right)}{2^{3/2}\left[\cosh(\tau-\tau')-\hat{n}\cdot\hat{n}'\right]^{3/2}}\\
G_{0}^{b}(r,r') & =\frac{1}{4\pi}\frac{1}{\sqrt{2}\sqrt{\cosh(\tau-\tau')-\hat{n}\cdot\hat{n}'}}
\end{align}
and the corresponding kernels are 
\begin{align}
D_{\ell}^{0}\left(\omega\right) & =\frac{1}{16\pi^{2}}\int_{r}e^{i\omega\tau}P_{\ell}\left(x\right)\begin{cases}
\frac{2}{2^{2}\left[\cosh(\tau)-\cos(\theta)\right]^{2}} & \mbox{GN}\\
\frac{1}{2\left[\cosh(\tau)-\cos(\theta)\right]} & \mbox{O}(N)
\end{cases}\nonumber \\
 & =\begin{cases}
\left(\ell^{2}+\omega^{2}\right)\mathcal{D}_{\ell-1}(\omega) & \mbox{GN}\\
\mathcal{D}_{\ell}(\omega) & \mbox{O}(N)
\end{cases}\label{eq:Dl0}
\end{align}
where 
\begin{equation}
\mathcal{D_{\ell}}(\omega)=\left|\frac{\Gamma\left(\frac{1+\ell+i\omega}{2}\right)}{4\Gamma\left(\frac{2+\ell+i\omega}{2}\right)}\right|^{2}.
\end{equation}

\subsubsection{Bosonic}

For non-zero topological charge, we turn to the spectral decomposition.
In the bosonic case, this is given by 
\begin{equation}
G_{Q_{B}}^{b}(r,r')=\dfrac{\left(1+x\right)^{Q_{B}}}{\left(4\pi\right)2^{Q_{B}}}e^{-2iQ_{B}\Theta}\sum_{\ell=Q_{B}}^{\infty}\frac{e^{-E_{Q_{B},\ell}^{b}|\tau-\tau'|}}{2E_{Q_{B},\ell}^{b}}(2\ell+1)P_{\ell-Q_{B}}^{\left(0,2Q_{B}\right)}\left(x\right),
\end{equation}
where $x=\hat{n}\cdot\hat{n}'$ , $e^{-2iQ_{B}\Theta}$ depends on
$\hat{n}$ and $\hat{n}'$ and 
\begin{align*}
E_{Q_{B},\ell}^{b} & =\sqrt{(\ell+1/2)^{2}+a_{Q_{B}}^{2}}.
\end{align*}
The kernel, after simplifications, becomes 
\begin{equation}
D_{\ell}^{Q_{B}}\left(\omega\right)=\frac{1}{16\pi}\sum_{\ell',\ell''=Q_{B}+1}\frac{(2\ell'+1)(2\ell''+1)\Bigl(E_{Q_{B},\ell'}^{b}+E_{Q_{B},\ell''}^{b}\Bigr)}{E_{Q_{B},\ell'}^{b}E_{Q_{B},\ell''}^{b}\left[\omega^{2}+\Bigl(E_{Q_{B},\ell'}^{b}+E_{Q_{B},\ell''}^{b}\Bigr)^{2}\right]}\begin{pmatrix}\ell & \ell' & \ell''\\
0 & -Q_{B}/2 & Q_{B}/2\\
0 & -Q_{B}/2 & Q_{B}/2
\end{pmatrix},\label{eq:Dl_ON}
\end{equation}
where the last factor is a product of two three-J symbols 
\begin{equation}
\left(\begin{array}{ccc}
\ell_{1} & \ell_{2} & \ell_{3}\\
m_{1} & m_{2} & m_{3}\\
m_{4} & m_{5} & m_{6}
\end{array}\right)\equiv\left(\begin{array}{ccc}
\ell_{1} & \ell_{2} & \ell_{3}\\
m_{1} & m_{2} & m_{3}
\end{array}\right)\left(\begin{array}{ccc}
\ell_{1} & \ell_{2} & \ell_{3}\\
m_{4} & m_{5} & m_{6}
\end{array}\right).
\end{equation}

\subsubsection{Fermionic}

In the case of the fermionic theory, obtaining the kernel in a simplified
form is much more involved given the spinor structure 
\begin{equation}
G_{Q_{B}}^{f}\left(r,r^{\prime}\right)=\frac{i}{2}\sum_{\ell=Q_{B}}^{\infty}e^{-E_{Q_{B};\ell}^{f}\left|\tau-\tau^{\prime}\right|}\sum_{m=-\ell}^{\ell-1}\left[S_{Q_{B},\ell-1,m}^{+}\quad S_{Q_{B},\ell,m}^{-}\right]\left(\text{sgn}\left(\tau-\tau^{\prime}\right)\mathbf{N}_{Q_{B},\ell}+\left(\begin{array}{cc}
0 & 1\\
-1 & 0
\end{array}\right)\right)\left[\begin{array}{c}
\left(S_{Q_{B},\ell-1,m}^{+}\right)^{\dagger}\\
\left(S_{Q_{B},\ell,m}^{-}\right)^{\dagger}
\end{array}\right],\label{eq:GF_monopole_harmonics_1}
\end{equation}
where 
\begin{align}
S_{Q_{B},\ell^{\prime},m^{\prime}}^{\pm} & =\begin{pmatrix}\pm\alpha_{\pm}Y_{Q_{B},\ell^{\prime},m^{\prime}}\\
\alpha_{\mp}Y_{Q_{B},\ell^{\prime},m^{\prime}+1}
\end{pmatrix},\quad\alpha_{\pm}=\sqrt{\frac{\ell^{\prime}+1/2\pm\left(m^{\prime}+1/2\right)}{2\ell^{\prime}+1}}\\
E_{Q_{B},\ell}^{f} & =\sqrt{\ell^{2}-Q_{B}^{2}}\\
\boldsymbol{N}_{Q_{B},\ell} & =-\frac{1}{\ell}\left(Q_{B}\tau_{z}+E_{Q_{B},\ell}^{f}\tau_{x}\right)
\end{align}
The fermionic Green's function is thus a $2\times2$ matrix whose
components are pairs of monopole harmonics 
\begin{equation}
G_{Q_{B}}^{f}\left(r,r^{\prime}\right)=\left(2\times2\text{ matrix}\right)\propto\sum_{\ell^{\prime}=Q_{B}}^{\infty}\sum_{m^{\prime}=-\ell^{\prime}+1}^{\ell^{\prime}}Y_{Q_{B},\ell^{\prime}+\delta\ell^{\prime},m^{\prime}+\delta m^{\prime}}\left(\hat{n}\right)Y_{Q_{B},\ell^{\prime}+\widetilde{\delta}\ell^{\prime},m^{\prime}+\widetilde{\delta}m^{\prime}}^{*}\left(\hat{n}^{\prime}\right),\quad\begin{cases}
\mathfrak{\delta\ell}^{\prime},\widetilde{\delta}\ell^{\prime}\in\{-1,0\}\\
\delta m^{\prime},\widetilde{\delta}m^{\prime}\in\{0,1\}
\end{cases}.\label{eq:GF_monopole_harmonics_2}
\end{equation}
In turn, the kernel momentum space coefficient takes the form 
\begin{equation}
D_{\ell}^{Q_{B}}\left(\omega\right)\sim\int_{x}\sum_{\ell',\ell''=Q_{B}}^{\infty}\sum_{m,m',m''}\text{product of 6 harmonics}.
\end{equation}
By using the following identities 
\begin{align}
Y_{Q_{B},\ell,m}^{*}\left(\hat{n}\right) & =\left(-1\right)^{Q_{B}+m}Y_{-Q_{B},\ell,-m}\left(\hat{n}\right)\\
Y_{Q_{B},\ell,m}\left(\hat{z}\right) & =\delta_{Q_{B},-m}\sqrt{\frac{2\ell+1}{4\pi}}
\end{align}
three harmonics are removed, and we are left with integrals of the
following form:

\begin{align}
\int d\hat{n}Y_{Q_{B},\ell,m}\left(\hat{n}\right)Y_{Q_{B}^{\prime},\ell^{\prime},m^{\prime}}\left(\hat{n}\right)Y_{Q_{B}^{\prime\prime},\ell^{\prime\prime},m^{\prime\prime}}\left(\hat{n}\right) & =\left(-1\right)^{\ell+\ell^{\prime}+\ell^{\prime\prime}}\sqrt{\frac{\left(2\ell+1\right)\left(2\ell^{\prime}+1\right)\left(2\ell^{\prime\prime}+1\right)}{4\pi}}\begin{pmatrix}\ell & \ell^{\prime} & \ell^{\prime\prime}\\
Q_{B} & Q_{B}^{\prime} & Q_{B}^{\prime\prime}\\
m & m' & m''
\end{pmatrix}.
\end{align}
The kernel coefficient obtained is

\begin{align}
D_{\ell}^{Q_{B}}(\omega) & =\mathfrak{D}_{\ell}^{Q_{B}}(\omega)+\sum_{\ell',\ell''=Q_{B}+1}^{\infty}\frac{\Bigl(-1\Bigr)^{\ell+\ell'+\ell''}\Bigl(E_{Q_{B},\ell'}^{f}+E_{Q_{B},\ell''}^{f}\Bigr)}{8\pi\left(\omega^{2}+\Bigl(E_{Q_{B},\ell'}^{f}+E_{Q_{B},\ell''}^{f}\Bigr)^{2}\right)}\nonumber \\
 & \times\left(\left[\left(\begin{array}{ccc}
\ell & \ell'-1 & \ell''-1\\
0 & -Q_{B} & Q_{B}\\
0 & Q_{B} & -Q_{B}
\end{array}\right)+\left(\begin{array}{ccc}
\ell & \ell' & \ell''\\
0 & -Q_{B} & Q_{B}\\
0 & Q_{B} & -Q_{B}
\end{array}\right)\right]2\left(Q_{B}^{2}+E_{Q_{B},\ell'}^{f}E_{Q_{B},\ell''}^{f}+\ell'\ell''\right)\right.\nonumber \\
 & \quad{}+\left[\left(\begin{array}{ccc}
\ell & \ell'-1 & \ell''\\
0 & -Q_{B} & Q_{B}\\
0 & Q_{B} & -Q_{B}
\end{array}\right)+\left(\begin{array}{ccc}
\ell & \ell' & \ell''-1\\
0 & -Q_{B} & Q_{B}\\
0 & Q_{B} & -Q_{B}
\end{array}\right)\right]2\left(Q_{B}^{2}+E_{Q_{B},\ell'}^{f}E_{Q_{B},\ell''}^{f}-\ell'\ell''\right)\nonumber \\
 & \quad{}-\left[\left(\begin{array}{ccc}
\ell & \ell' & \ell''\\
0 & -Q_{B}-1 & Q_{B}+1\\
0 & Q_{B} & -Q_{B}
\end{array}\right)s_{\ell'+1}^{+}s_{\ell''+1}^{+}+\left(\begin{array}{ccc}
\ell & \ell'-1 & \ell''-1\\
0 & -Q_{B}-1 & Q_{B}+1\\
0 & Q_{B} & -Q_{B}
\end{array}\right)s_{\ell'-1}^{-}s_{\ell''-1}^{-}\right]\left(s_{\ell'}^{-}s_{\ell''}^{-}+s_{\ell'}^{+}s_{\ell''}^{+}\right)\nonumber \\
 & \quad{}+\left(\begin{array}{ccc}
\ell & \ell'-1 & \ell''\\
0 & -Q_{B}-1 & Q_{B}+1\\
0 & Q_{B} & -Q_{B}
\end{array}\right)s_{\ell'-1}^{-}s_{\ell''+1}^{+}\left(s_{\ell''}^{-}s_{\ell'}^{+}-s_{\ell'}^{-}s_{\ell''}^{+}\right)\nonumber \\
 & \quad{}-\left(\begin{array}{ccc}
\ell & \ell' & \ell''\\
0 & 1-Q_{B} & Q_{B}-1\\
0 & Q_{B} & -Q_{B}
\end{array}\right)s_{\ell'}^{-}s_{\ell'+1}^{-}s_{\ell''}^{-}s_{\ell''+1}^{-}-\left(\begin{array}{ccc}
\ell & \ell'-1 & \ell''-1\\
0 & 1-Q_{B} & Q_{B}-1\\
0 & Q_{B} & -Q_{B}
\end{array}\right)s_{\ell'-1}^{+}s_{\ell'}^{+}s_{\ell''-1}^{+}s_{\ell''}^{+}\nonumber \\
 & \quad{}+\left(\begin{array}{ccc}
\ell & \ell'-1 & \ell''\\
0 & 1-Q_{B} & Q_{B}-1\\
0 & Q_{B} & -Q_{B}
\end{array}\right)s_{\ell''}^{-}s_{\ell'-1}^{+}s_{\ell'}^{+}s_{\ell''+1}^{-}-\left(\begin{array}{ccc}
\ell & \ell'-1 & \ell''\\
0 & 1-Q_{B} & Q_{B}-1\\
0 & Q_{B} & -Q_{B}
\end{array}\right)s_{\ell'}^{-}s_{\ell'-1}^{+}s_{\ell''}^{+}s_{\ell''+1}^{-}\nonumber \\
 & \quad{}-\left(\begin{array}{ccc}
\ell & \ell' & \ell''\\
0 & 1-Q_{B} & Q_{B}-1\\
0 & Q_{B} & -Q_{B}
\end{array}\right)s_{\ell'+1}^{-}s_{\ell'}^{+}s_{\ell''}^{+}s_{\ell''+1}^{-}+\left(\begin{array}{ccc}
\ell & \ell' & \ell''-1\\
0 & -Q_{B}-1 & Q_{B}+1\\
0 & Q_{B} & -Q_{B}
\end{array}\right)s_{\ell''-1}^{-}s_{\ell''}^{-}s_{\ell'}^{+}s_{\ell'+1}^{+}\nonumber \\
 & \quad{}-\left(\begin{array}{ccc}
\ell & \ell'-1 & \ell''-1\\
0 & 1-Q_{B} & Q_{B}-1\\
0 & Q_{B} & -Q_{B}
\end{array}\right)s_{\ell'}^{-}s_{\ell''}^{-}s_{\ell'-1}^{+}s_{\ell''-1}^{+}-\left(\begin{array}{ccc}
\ell & \ell' & \ell''-1\\
0 & 1-Q_{B} & Q_{B}-1\\
0 & Q_{B} & -Q_{B}
\end{array}\right)s_{\ell'+1}^{-}s_{\ell''}^{-}s_{\ell'}^{+}s_{\ell''-1}^{+}\nonumber \\
 & \quad{}\left.-\left(\begin{array}{ccc}
\ell & \ell' & \ell''-1\\
0 & -Q_{B}-1 & Q_{B}+1\\
0 & Q_{B} & -Q_{B}
\end{array}\right)s_{\ell'}^{-}s_{\ell''-1}^{-}s_{\ell'+1}^{+}s_{\ell''}^{+}+\left(\begin{array}{ccc}
\ell & \ell' & \ell''-1\\
0 & 1-Q_{B} & Q_{B}-1\\
0 & Q_{B} & -Q_{B}
\end{array}\right)s_{\ell'}^{-}s_{\ell'+1}^{-}s_{\ell''-1}^{+}s_{\ell''}^{+}\Biggr)\right),\label{eq:Dl_GN}
\end{align}
where $s_{\ell}^{\pm}=\sqrt{\ell\pm Q_{B}}$ and where $\mathfrak{D}_{\ell}^{Q_{B}}(\omega)$,
the contribution when one of the two angular momentum has the minimal
value $Q_{B}$, is given by 
\begin{align}
\mathfrak{D}_{\ell}^{Q_{B}}(\omega) & =\sum_{\ell'=Q_{B}+1}^{\infty}\frac{-(-1)^{Q_{B}+\ell'+\ell}E_{Q_{B},\ell'}}{8\pi\left(E_{Q_{B},\ell'}^{2}+\omega^{2}\right)}\left(2Q_{B}\left[-\left(s_{\ell''}^{-}\right)^{2}\left(\begin{array}{ccc}
\ell & \ell'-1 & Q_{B}\\
0 & Q_{B} & -Q_{B}\\
0 & -Q_{B} & Q_{B}
\end{array}\right)+\left(s_{\ell''}^{+}\right)^{2}\left(\begin{array}{ccc}
\ell & \ell' & Q_{B}\\
0 & Q_{B} & -Q_{B}\\
0 & -Q_{B} & Q_{B}
\end{array}\right)\right]\right.\nonumber \\
 & \left.-\sqrt{2Q_{B}}\left[s_{\ell'}^{-}s_{\ell'-1}^{+}\left(\begin{array}{ccc}
\ell & \ell'-1 & Q_{B}\\
0 & Q_{B} & -Q_{B}\\
0 & 1-Q_{B} & Q_{B}-1
\end{array}\right)+s_{\ell'+1}^{-}s_{\ell'}^{+}\left(\begin{array}{ccc}
\ell & \ell' & Q_{B}\\
0 & Q_{B} & -Q_{B}\\
0 & 1-Q_{B} & Q_{B}-1
\end{array}\right)\right]\right).
\end{align}

\subsubsection{Regularizing the coefficient}

The kernel momentum space coefficient $D_{\ell}^{Q_{B}}(\omega)$
(see Eq.~\eqref{eq:Dl_ON} for O$(N)$ model and Eq.~\eqref{eq:Dl_GN}
for $\mbox{GN}$ model) is needed to compute the monopole anomalous dimension
\eqref{eq:anomalous_dimension}. Computing $D_{\ell}(\omega)$ requires
performing two infinite sums 
\begin{equation}
D_{\ell}^{Q_{B}}(\omega)\equiv\sum_{\ell^{\prime},\ell^{\prime\prime}}^{\infty}d_{\ell,\ell^{\prime},\ell^{\prime\prime}}^{Q_{B}}\left(\omega\right).\label{eq:Dl_sum}
\end{equation}
There is a double sum, but the argument vanishes unless $\ell,\ell'$
and $\ell''$ satisfy the triangle equality. This constraint effectively
removes the divergence of one integral. The remaining effect can be
considered by fixing $\ell$ and studying the argument in the sum
at large $\ell'$. 
\begin{equation}
\lim_{\ell^{\prime}\to\infty}\sum_{\ell^{\prime\prime}=Q_{B}+1}^{\infty}d_{\ell,\ell^{\prime},\ell^{\prime\prime}}^{Q_{B}}\left(\omega\right)=\xi=\begin{cases}
-\frac{1}{2\pi} & \mbox{GN}\\
0 & \mbox{O}(N)
\end{cases}.
\end{equation}
The constant contribution in this limit in the $\mbox{GN}$ model leads
to a divergence which must be regularized. 
\begin{equation}
\zeta(0,Q_{B}+1)\xi+\sum_{\ell^{\prime}=Q_{B}+1}^{\infty}\biggl[-\xi\sum_{\ell^{\prime\prime}=Q_{B}+1}^{\infty}d_{\ell,\ell^{\prime},\ell^{\prime\prime}}^{Q_{B}}\left(\omega\right)\biggr]\equiv\zeta(0,Q_{B}+1)\xi+\sum_{\ell^{\prime}=Q_{B}+1}^{\infty}\tilde{d}_{\ell,\ell'}^{Q_{B}}(\omega).
\end{equation}
Subleading terms in both cases scale as $\ell'^{-2}$ and do not pose
a problem for convergence.

To speed up the numerical computation, the summand $\tilde{d}_{\ell,\ell'}^{Q_{B}}(\omega)$
can be expanded analytically as inverse powers of $\ell'$: 
\begin{equation}
\tilde{d}_{\ell,\ell'}^{Q_{B}}(\omega)=\begin{cases}
\frac{\left(\ell(\ell+1)-4Q_{B}^{2}+2\omega^{2}\right)}{16\pi\ell^{\prime2}}+\frac{\left(7\left[\ell(\ell+1)\right]^{2}+\ell(\ell+1)\left(24Q_{B}^{2}+8\omega^{2}-2\right)-8\left(-6Q_{B}^{2}\omega^{2}+6Q_{B}^{4}+\omega^{4}\right)\right)}{256\pi\ell^{\prime4}}+\dots & \mbox{GN}\\
\frac{1}{8\pi\ell^{\prime2}}-\frac{1}{8\pi\ell^{\prime3}}+\frac{\left(-6(a_{Q_B}^2)-\omega^{2}+2\ell(\ell+1)+3\right)}{32\pi\ell^{\prime4}}+\frac{\left(6(a_{Q_B}^2)+\omega^{2}-2\ell(\ell+1)-1\right)}{16\pi\ell^{\prime5}}+\dots & \mbox{O}(N)
\end{cases}\label{eq:remainder_expansion}
\end{equation}
The numerical sum can then be stopped to a relatively low cut-off
$\ell'_{c}$, and the rest of the sum is handled analytically $\sum_{\ell'=\ell'_{c}+1}^{\infty}\left(\ell'\right)^{-p}=\zeta(p,\ell'_{c}+1)$,
and 
\begin{equation}
\sum_{\ell'=Q_{B}+1}^{\infty}\tilde{d}_{\ell,\ell'}^{Q_{B}}(\omega)\approx\sum_{\ell'=Q_{B}+1}^{\ell'_{c}}\tilde{d}_{\ell,\ell'}^{Q_{B}}(\omega)+\sum_{p=2}c_{\ell}^{(p)}(\omega)\zeta(p,\ell'_{c}+1),
\end{equation}
where the coefficients $c_{\ell}^{(p)}(\omega)$ are read off from
Eq.~\eqref{eq:remainder_expansion}. We used $\ell'_{c}=150$ and
$p=15$. This allows to obtain the regularized $D_{\ell}^{Q_{B}}(\omega)$~\eqref{eq:Dl_sum},
which is inserted in the monopole anomalous dimension ~\eqref{eq:anomalous_dimension}
along with the zero charge kernel~\eqref{eq:Dl0}. The remaining
sum on $\ell$ and integral on $\omega$ are computed up to a relativistic
cut-off 
\begin{equation}
\ell\left(\ell+1\right)+\omega^{2}\leq L\left(L+1\right).
\end{equation}
For the $\mbox{GN}$ model, we used $L=35+\text{Round}(Q_{B})$. For the
$\mbox{O}(N)$ model we use $L=30$ along with a UV expansion in $\ell,\omega$
to the obtain the region $L\in]30,\infty[$ as in Ref.~\citep{dyer_scaling_2015}.

\twocolumngrid

\bibliographystyle{apsrev}
\bibliography{ref2}

\end{document}